\documentclass{sig-alternate-10pt}
\usepackage{listings}
\begin{document}
\title{SmartNight: Turning Off the Lights on Android}
\numberofauthors{1}
\author{
\alignauthor
Andrew Banman\\
       \affaddr{University of Minnesota}\\
       \affaddr{100 Church St. S.E.}\\
       \affaddr{Minneapolis, Minnesota}\\
       \email{banma001@umn.edu}
}

\maketitle
\begin{abstract}
Smartphone users benefit from content with dark color schemes: increasingly common OLED displays are more power efficient the darker the display, and many users prefer a dark display for night time use. Despite these benefits, many applications and the majority of web content are drawn with white backgrounds. There are many partial solutions to darken the displayed content, but none work in all situations. Enter SmartNight, a content-aware solution to dynamically darken content on Android. By trading off content fidelity, Android with SmartNight displays content with nearly 90\% lower average picture level. It is implemented in the Android framework, and requires no external support. It seamlessly incorporates existing solutions, making it a bridge between the state-of-the-art and future solutions.
\end{abstract}

\section{Introduction}

Dark color schemes on mobile phones are beneficial to users in at least three ways. First, organic light-emitting diode (OLED) displays consume less power the darker the display. Each pixel in an OLED display is individually lit, meaning a pixel presenting black color in theory draws no power. A smartphone's display is the biggest individual energy consumer on the device (for uses other than phone calls) \cite{carroll} so the improved efficiency of OLED is highly desirable. OLED displays are now commonplace, such as in the new Samsung Galaxy S10 smartphone \cite{s10shootout}. Dark content gives the user the most energy benefit from this technology.

Second, bright screens during nighttime smartphone use cause mild health problems. Medical research has shown white light suppresses an individual's melatonin levels, interrupting their circadian rhythm and leading to irregular sleep. Some studies and preliminary measurements have even linked inhibited melatonin with diabetes, heart disease, and some types of cancer. It has also been shown that blue light has a larger effect on melatonin levels than red or green light, which explains the recent popularity of display-warming solutions \cite{flux}, where each pixel is shifted towards red while preserving relative color differences. \cite{harvardhealth}

These health concerns are related to the general discomfort caused by bright lights in a dark room. Many professional programmers choose light-on-dark color schemes to reduce eye strain. Others simply prefer the appearance of these themes. Whatever the motivation, there appears to be a large demand for more dark content.

\subsection{Related Work}

The most familiar solutions to the white screen problem are application ``dark'' (or ``night'') modes. Typically these modes swap the default UI color scheme with a dark grey or black background and light grey or white text. Dark modes are gaining popularity, and it would be very interesting to quantify this trend with a survey of applications providing dark modes. This trend towards dark schemes is exemplified by Google's decision to include a global dark mode in Android Q. According to recent rumors, the user can set the UI globally to a dark theme while automatically switching applications to their internal dark modes. \cite{androidQ_rumor}

In an ideal world all application developers would ship a dark mode, but even this hypothetical does not account for the fact that the majority of web content is dominated by white pixels. One study found that 80\% of website content is white. \cite{laaperi} Web browsing has traditionally been problematic on mobile phones for many reasons, but its too important a use case to ignore. This specific challenge is addressed by Chameleon browser \cite{chameleon}, a custom browser solution that uses a proxy server to dynamically change the color of websites to produce darker content. Chameleon gives good results, but it's eventual implementation seems unlikely since it requires extensive proxy support. A local solution that doesn't depend on external services is more likely to be used.

In this paper I explore a global, content-aware solution that dynamically transforms bright content into dark content. The goal is to balance the tradeoff between content fidelity and dark screens, with a bias towards the latter. My solution, called \emph{SmartNight}, fills the gaps between existing dark-mode solutions with minimal adoption barriers. It is implemented in Android 9.0.

The rest of the paper proceeds as follows: \S\ref{sec:graphics} covers the necessary background on Android graphics frameworks, \S\ref{sec:design} details SmartNight's implementation, I evaluate SmartNight's effectiveness in \S\ref{sec:results}, and finally there is discussion and recommendations for future work in \S\ref{sec:conclusion}.

\subsection{Android Graphics Stack}
\label{sec:graphics}

Application views are composed of multiple windows typically three at any time: the status bar, application UI, and navigation bar. Each window is backed by an independent \emph{Layer} provided by the \emph{SurfaceFlinger} service. Note that there are different names for the drawing surface in different contexts: window in the context of the WindowManager or layer in SurfaceFlinger's parlance. SurfaceFlinger is responsible for creating the independent Layers, compositing them, and sending them to the display hardware for the final rendering.

Each Layer holds a \emph{BufferQueue} that synchronously shares graphical buffers between the producer (i.e. the application), and its consumers (e.g. SurfaceFlinger). Importantly, buffers are shared by reference via a synchronous latching mechanism; they are never copied. BufferQueue producers are implemented by the \emph{Surface} class, part of the public java API. Most Surfaces are eventually consumed by SurfaceFlinger, but others are rendered directly by the application.

\section{SmartNight Design}
\label{sec:design}
At a high level, SmartNight consists of two major steps: an augmented SurfaceFlinger 1) performs some straightforward content analysis to decide whether or not to transform each visible Layer, and 2) transforms the color the Layers identified in step 1. The transformation is accomplished by applying a color transformation matrix. These steps are covered in \S\ref{sec:content} and \S\ref{sec:colortransform}, respectively. As we will see, there is a large design space both in the analysis method and in the color transformation itself.

The SurfaceFlinger service is a good candidate to host this feature because almost every rendered object passes through it and its client-server model permits asynchronous analysis. These two features gives SmartNight a very general, low-level approach with small impact on display latency.

\subsection{Content Analysis}
\label{sec:content}
SmartNight's goal is to transform bright-dominant screens to dark-dominant screens, while preserving those that are already dark-dominant. To do this it must determine what are the dominant colors of each currently visible Layer. Luckily, SurfaceFlinger can do this analysis more-or-less asynchronously with respect to the actual rendering. When SurfaceFlinger receives an asynchronous invalidation message from one of its Surface producers, it inspects the Layer for changes. If the Layer has been latched, we infer the content was updated and a refresh is needed. SmartNight inserts the content analysis step at this point in the inspection.

Content analysis is done in three phases: sampling the Layer's graphical buffer, computing the approximate overall luminance of the image, and comparing the luminance against threshold values to determine whether the content is too bright and thus should be modified. Additional steps can (and should) be added to improve the judgement. For example, video buffers should be identified and judged not be transformed, so as not toe disrupt playback.

\subsubsection{Sampling}
This initial version of SmartNight uses a very simple sampling method to determine whether the image is comprised of more bright or dark pixels. Roughly 2500 pixels are sampled by incrementing through vertical and horizontal strides. The horizontal stride is periodically offset to avoid sampling along vertical lines through the 2-dimensional image. This offset prevents oversampling vertically aligned colors, such as a narrow vertical bar or justified text, and improves the accuracy of smaller sample sizes.

Each pixel is extracted according to the buffer's pixel format, as specified in the containing Layer. The pixel format specifies the byte layout of the red, green, blue, and sometimes alpha values. In this version the alpha channel is discarded, but it may prove useful to other methods. I also record the maximum values of each channel for later comparison.

An alternative method could be to search for the background color instead of the overall dominant color. Assuming the background color is constant (or nearly so), rather than sample scattered individual pixels look at contiguous groups of pixels. If a each pixel in the group has identical color then it is very likely it matches the background color. This can be part of an optimization: keeping a similar stride pattern as before, inspect each pixel in a cache-aligned block and only include it in the overall count if its neighbors have the same (or very nearly the same) RGB values. In this manner you can exclude pixels making up foreground images in trivial additional time.

\subsubsection{Luminance}
There are various formula for computing luminance from RGB color channels. In this implementation I use a simplified formula, that nevertheless accounts for the different brightness contrubtions of each RGB channel: $l = 3r + b + 4g$. The simplification is suitable because 1) computation should be as fast as possible, and 2) we don't need precise results to make a good judgement.

\subsubsection{Threshold}
Once the average luminance of the layer is computed we must decide whether to perform the inversion. Setting this threshold too high will permit more white displays to be shown, while setting it too close to the center will be less stable. I arrived, through trial and error, at 60\% of full white as the floor for a \emph{bright} pixel and 40\% as the ceiling for a \emph{dark} pixel. If the bright pixels outnumber the dark then a flag is set to direct SurfaceFlinger to invert the entire Layer. Pixels within the range are not counted.

\subsection{Selective Color Transform}
\label{sec:colortransform}

As discussed, almost every rendered object passes through the SurfaceFlinger service, and each of these is backed by a Layer. During the composition step, SurfaceFlinger renders each visible Layer sequentially (according to its z-position) into earlier-computed cropped dimensions. It turns out one can set the color transform for the renderer at the time SurfaceFlinger draws each layer. This permits the simple logic: \emph{if the layer should be transformed, then set the renderer's color matrix and draw, else clear the render's color matrix and draw.} But is this an efficient method?

Since Android 3.0, all images associated with and Application's windows are rendered using hardware acceleration. In other words, the GPU is the default renderer for all applications. GPUs are extremely efficient at performing vector computations, such as multiplying a color transformation matrix. Thus, by default we can swap color matrices virtually for free. The contradicting case occurs when applications explicitly opt to use CPU rendering, but rendering of this kind escapes SurfaceFlinger anyway, as discussed in \S\ref{sec:graphics}.

Furthermore, the result of the content analysis could select an specific color transformation. There's nothing preventing this system from adopting a per-layer color matrix. For example, images not as severely dominated by bright pixels could receive a milder color transformation thank full color inversion. The result could improve the darkening and/or the content fidelity.

While the overall approach is very general, certain objects -- such as the UI keyboard -- escape transformation. In fact, anything drawn directly by the application via OpenGL will not be consumed by SurfaceFlinger and thus circumvents SmartNight. This is the most visible drawback of the current implementation. An older approach instead modified the OpenGL renderer to globally apply color transformations, but it's not clear how to conduct the content-analysis in that case. I wasn't able to locate the original 2010 patch, and its beyond my knowledge whether an updated version would be feasible nearly 10 years later. \cite{sharkey}

A fully-implemented version of SmartNight would include a UI module, allowing the user to choose between different color transformations as well as to toggle the feature on and off. For expediency's sake, this version of SmartNight has no UI. Rather, it is always on and switches between no color transformation, and the present globally-defined transformation. In native Android the only global transform typically used is the color inversion provided by an accessibility mode. Luckily, color inversion is a good starting point for a global dark mode.

\section{Results}
\label{sec:results}
There are two main dimensions on which to evaluate SmartNight and its competitors. The first dimension is display power reduction. One would like to know by how much SmartNight reduces battery consumption for OLED devices. In absence of real power measurements, one can measure how dark on average the same images rendered with SmartNight enabled versus the competition. A common OLED display metric is the average picture level (APL). I compare SmartNight to the default display as well as to a red-shift strategy in \S\ref{sec:apl}.

The second dimension, quality of experience (QoE), also needs evaluation. Eventually a user study must be conducted to establish ground truth for preferences. That effort is premature in the face of some necessary ironing-out work. What suffices is a qualitative discussion of the SmartNight experience along with preliminary quality measurements in \S\ref{sec:qoe}

\subsubsection{Darkening Performance}
\label{sec:apl}

Each OLED device has its own power characteristic, but they all share the property that power increases the closer the display is to full white. APL measures the overall distance from black (0) to full white (1). Thus, APL serves as a proxy measure of the power required to display an image. I compare the APL for common Android screens and for popular websites in Figures 1 and 2. For each image I calculate the APL of the default display, SmartNight, and red-shifted versions of each where only the red pixels are counted. This is an extreme flavor of red-shift that gives a low bound on the APL for such display-warming schemes.

\begin{figure}
    \label{fig:aplAndroid}
    \includegraphics[width=0.5\textwidth]{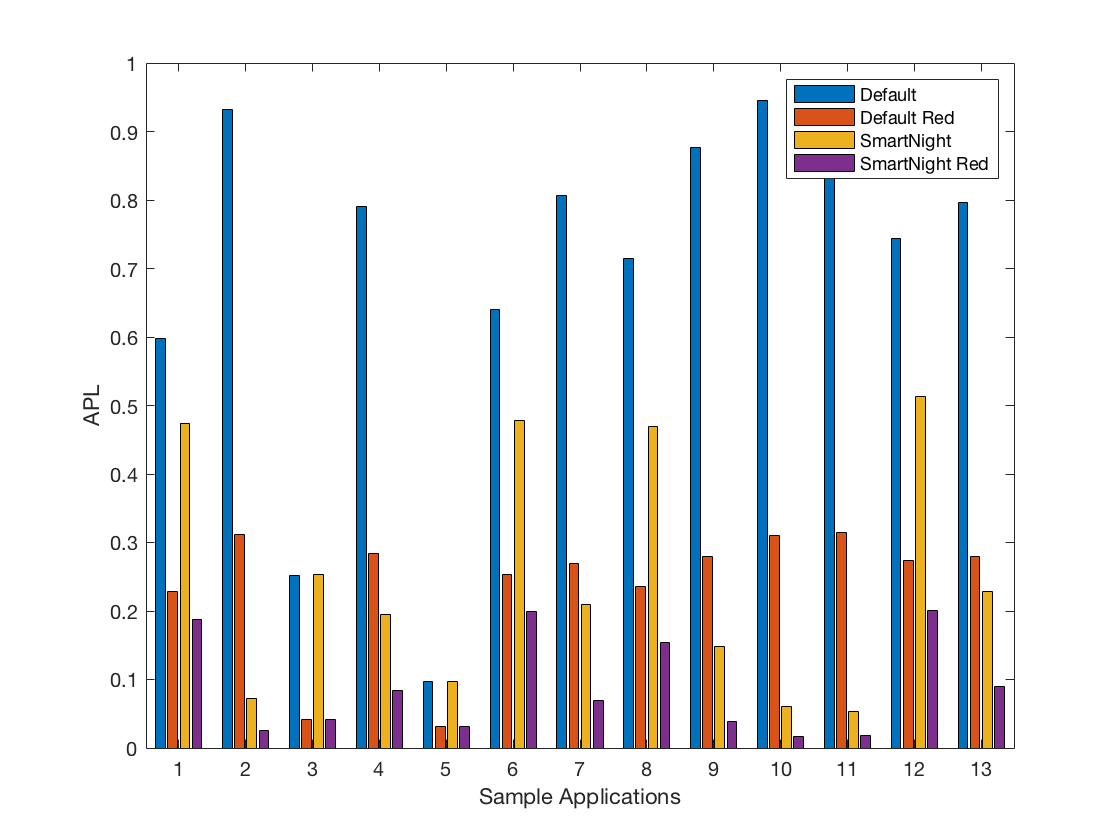}
    \caption{Average Picture Level of pre-installed applications and common android screens. Larger APL values indicate an image closer to full white.}
\end{figure}

\begin{figure}
    \label{fig:aplWeb}
    \includegraphics[width=0.5\textwidth]{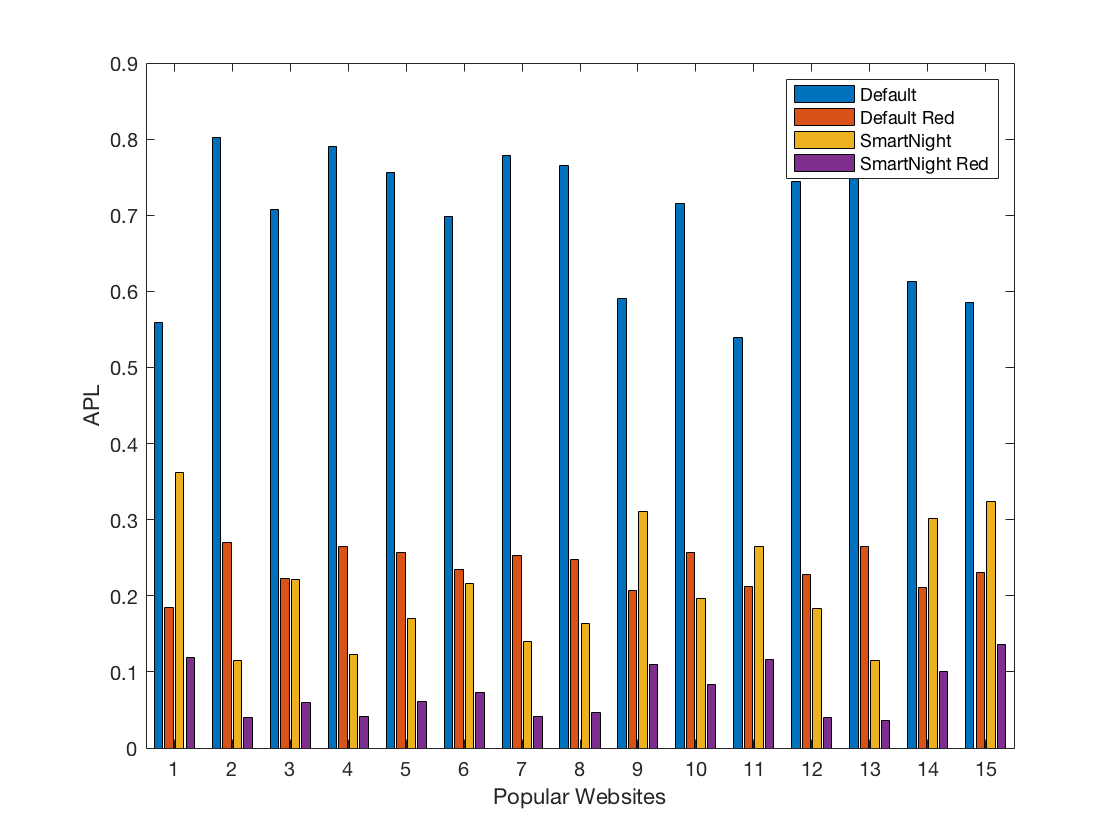}
    \caption{Average Picture Level of the top 15 most popular webpages. Larger APL values indicate an image closer to full white.\cite{alexa}}
\end{figure}

In every case, the default display has the highest APL (as expected) and SmartNight had lower APL, a mean reduction of 72\%. The mean default APL was 0.75, corroborating the result from \cite{laaperi} that smartphone screens are dominated by bright pixels. SmartNight also lowered APL by more than the default red-shift transformation in 17 cases, while there were 11 cases where red-shift had lower APL than SmartNight. These images were near the bright content cutoff, meaning the inversion has a smaller affect on APL. However, SmartNight $\times$ red-shift did better than red-shift in 92\% of cases, was equivalent in the remaining cases, and reduced APL by 89\% relative to default. Figure 3 shows the darkening of each scheme relative to the default display, and conclusively shows SmartNight's superior darkening effect, especially when combined with red-shift.

\begin{figure}
    \label{fig:darkening}
    \includegraphics[width=0.5\textwidth]{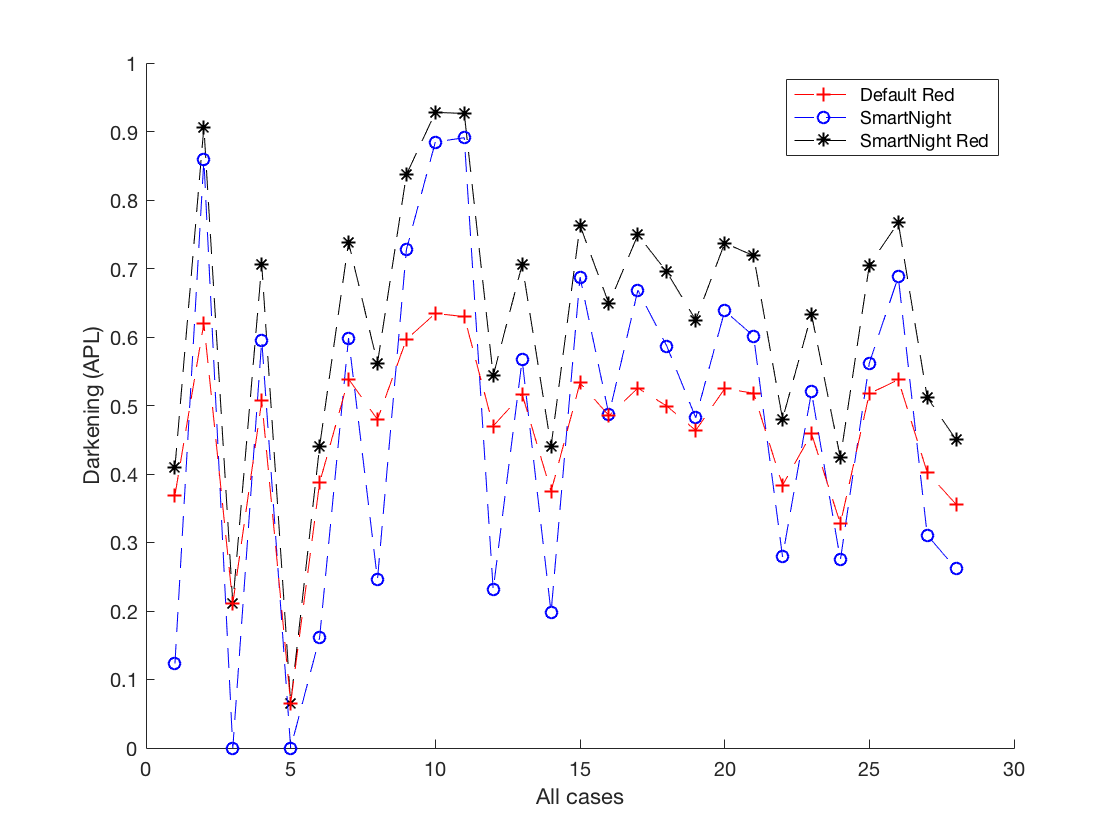}
    \caption{Average Picture Level reduction relative to the default image of common android screens and popular webpages. Larger values indicate larger APL reductions, i.e. a darker overall affect.}
\end{figure}

\subsubsection{Quality of Experience}
\label{sec:qoe}

The QoE is further divided into quality based on latency -- i.e. how smooth the display appears -- and the consistency of the color representation.

\subsubsection{Latency}
SmartNight was evaluated on Android emulation software shipped with the Android Open Source Project (AOSP) source code. Various graphics indicators were measured with the android debugger (adb) tool, specifically with the \texttt{dumpsys} sub command.

As discussed previously, there should be no degradation in video latency because the bulk of the work is done asynchronously with layer rendering and applying color transformations is very cheap. There was no perceived slowdown or framerate drop. Frames per second did not drop below the maximum 60fps throughout video playback on the VLC application.

However, SmartNight performed poorly in terms of the number of ``janky'' frames (those deemed too slow) and rendering speed distribution. Table 1 summarizes these results. The default color inversion feature also incurred a small performance hit. This is most likely due to the added cost of applying the color transformation to each layer. These results need to be corroborated by real hardware measurements, but they do show what appears to be real performance loss. However, extra time spent rendering frames will not impact power significantly since CPU is not a major consumer. Furthermore, the rendering speed is not the final word on display health. More study is needed to assess whether the performance is in acceptable bounds relative to the benefits of darkening.

\begin{table}
    \label{framedata}
    \begin{tabular}{l c c c}
        \hline
        & SmartNight & Default & Inverted \\
        \hline
        \hline
        Frames rendered:  & 1048    & 1189    & 1079     \\
        Janky frames      & 79.48\% & 27.17\% & 33.92\%  \\
        50th percentile:  & 19ms    & 9ms     & 10ms     \\
        90th percentile:  & 27ms    & 22ms    & 24ms     \\
        95th percentile:  & 34ms    & 25ms    & 28ms     \\
        99th percentile:  & 61ms    & 44ms    & 57ms     \\
        \hline
    \end{tabular}
    \caption{Framerate statistics gathered with adb while navigating the VLC application by hand. About 1000 frames were sampled in each case over the course of similar input sequence. Janky frames were identified as too slow. Various percentiles in the distribution of rendering speeds are also shown. The default display and the default color inversion feature were measured as well as SmartNight}
\end{table}

A final note on latency measurements: it seems like a contradiction that framerate is not affected while the average rendering time increased for SmartNight. This may be due to peculiarities with how the frames per second are counted in the emulation environment. Another peculiarity I observed while testing was that video frames did not seem to be counted where I expected them under the VLC package. It could be they were not factored into the reported framerate by adb. Nevertheless, I observed no video slowdown or jitter.

\subsubsection{Visual Consistency}
SmartNight has some problems with visual artifacts. The first one a user would notice is white flickering during some transitional animations, such as minimizing and application or swiping from the right. These are probably caused by incorrect content analysis placement. In this implementation, the content analysis is only done when SurfaceFlinger receives an invalidate message from a client producer. I suspect this code path is not exercised for the first drawn frame of a layer. A likely fix is to duplicate the content analysis at layer creation. In fact, this may be a workaround for the latency problem: rather than re-analyze a layer every time it changes, it may be sufficient to analyze the above the fold content and infer that the rest follows a similar color scheme. More tinkering is called for.

Video also remains a problem for SmartNight. When viewing video in portrait mode there are often large black bars above and below the video -- as a consequence of maintaining wide aspect ratios -- and in that case most frames will be categorized as dark and therefore won't be transformed. Unfortunately in the general case it's more than likely a bright frame from the video will trip the transformation. This looks like a flickering video, alternating between inverted colors and real colors.

I envision various potential solutions to the video problem. First, one could identify the type of producer for each Layer's BufferQueue; if the buffer came from a video stream then we can skip the content analysis step and decide not to transform the colors. In the best case, the identification would follow from metadata belonging to the BufferQueue, Layer, or some other encapsulating object. Initial experiments along this line failed, but I wouldn't rule out the option. The Layer's window type was tested (specifically \texttt{TYPE\_APPLICATION\_MEDIA}), as well as GraphicBuffer's usage flags. These were early shots in the dark but, with more experience and investigation, I have a hunch the answer is hiding somewhere in the SurfaceFlinger stack. Alternatively, perhaps the necessary information could be added to the Layer at the time of creation in another patch. Failing the simple solution, one could experimentally guess video content by taking buffer update timings -- presumably frequently updated layers are displaying video playback. Finally, the content analysis algorithm could be smart enough to ignore most video content. Previously I suggested an algorithm for determining the background color by looking for contiguous, identical pixels. With clever shape selection of these regions one could rule out all manner of natural images typically seen in video.

\section{Conclusions and Future Work}
\label{sec:conclusion}

SmartNight is a novel step towards global night mode on Android smartphones. It works well with existing night-optimized content, like dark websites and application-specific night modes. Initial measurements show a simple content-aware color inverter algorithm reduces APL by 72\%, and that this can be pushed as high as 89\% with color warming transformation. These gains come at no perceived framerate drop, although there are some artifacts that negatively effect the QoE.

There are a number of areas for improvement that make this an interesting project to continue. More work needs to be done to make the experience more smooth; in particular the color flickering problem must be solved. Research should be done on how to incorporate non-SurfaceFlinger-rendered buffers, and videos should be more robustly excluded from the darkening color transformation.

If and when this work is complete, a thorough technical evaluation is needed to definitely measure any energy savings on OLED devices. Next, a user study must quanitativley assess the QoE, which ultimately comes down to a tradeoff between content fidelity and night time comfort. I predict that a significant portion of users will appreciate the benefits of SmartNight enough to forgive the occasional inverted picture.

\bibliographystyle{abbrv}
\bibliography{report}

%
%

\end{document}